\documentclass{osa-article}

\journal{boe}

\articletype{Research Article}
\usepackage{slashbox}
\usepackage{multirow}
\begin{document}

\title{Multi-scale GCN-assisted two-stage network for joint segmentation of retinal layers and disc in peripapillary OCT images}
    
\author{Jiaxuan Li,\authormark{1} Peiyao Jin,\authormark{2,3} Jianfeng Zhu,\authormark{2} Haidong Zou,\authormark{2,3} Xun Xu,\authormark{2,3} Min Tang,\authormark{3} Minwen Zhou,\authormark{3}  Yu Gan,\authormark{4}  Jiangnan He\authormark{2,*} Yuye Ling\authormark{1,*} and Yikai Su\authormark{5}}

\address{\authormark{1}John Hopcroft Center for Computer Science, Shanghai Jiao Tong University, Shanghai 200240, China\\
\authormark{2}Department of Preventative Ophthalmology, Shanghai Eye Disease Prevention and Treatment Center, Shanghai Eye Hospital, Shanghai 200040, China\\
\authormark{3}Department of Ophthalmology, Shanghai General Hospital, Shanghai Jiao Tong University, Shanghai 200080, China\\
\authormark{4}Department of Electrical and Computer Engineering, The University of Alabama, AL 35487, USA\\
\authormark{5}State Key Lab of Advanced Optical Communication Systems and Networks, Department of Electronic Engineering, Shanghai Jiao Tong University, Shanghai 200240, China}
\email{\authormark{*}yuye.ling@sjtu.edu.cn and hejiangnan85@126.com } 



\begin{abstract}
An accurate and automated tissue segmentation algorithm for retinal optical coherence tomography (OCT) images is crucial for the diagnosis of glaucoma. However, due to the presence of the optic disc, the anatomical structure of the peripapillary region of the retina is complicated and is challenging for segmentation. To address this issue, we developed a novel graph convolutional network (GCN)-assisted two-stage framework to simultaneously label the nine retinal layers and the optic disc. Specifically, a multi-scale global reasoning module is inserted between the encoder and decoder of a U-shape neural network to exploit anatomical prior knowledge and perform spatial reasoning. We conducted experiments on human peripapillary retinal OCT images. The Dice score of the proposed segmentation network is 0.820 ± 0.001 and the pixel accuracy is 0.830 ± 0.002, both of which outperform those from other state-of-the-art techniques.
\end{abstract}

\section{Introduction}
Glaucoma is the leading cause of irreversible blindness globally, which affects approximately 64.3 million individuals worldwide \cite{RN145}. In China, around 13.12 million people were affected by glaucoma in 2015, and the number is projected to reach 25.16 million by 2050 \cite{RN144}. This will cause a heavy burden on public health. Currently, the most effective way to prevent glaucoma- related vision loss is early diagnosis and early intervention. In particular, detecting small morphological changes of retinal layers, such as the thinning of the retinal nerve fiber layer (RNFL) and ganglion cell layer (GCL), has critical value on the precise diagnosis of glaucoma  \cite{RN143}.

Optical coherence tomography (OCT) as a non-invasive three-dimensional imaging modality is commonly used in eye clinics for retinal inspection. Due to the micro-meter-level axial resolution, it provides a unique capability to directly visualize the stratified structure of the retina and access their corresponding thicknesses. OCT-derived thickness of peripapillary RNFL is a common indicator in early-stage glaucoma diagnosis \cite{RN146}. Therefore, a precise tissue segmentation of retinal OCT images becomes a critical step towards successful early diagnosis of glaucoma.

However, manual segmentation is time-consuming and laborious, while an accurate automated algorithm is desirable by both clinicians and researchers. Numerous automated retinal OCT segmentation techniques have been proposed in the past decades \cite{RN122,RN129,RN120,RN130,RN121,RN123,RN125,RN113,RN128,RN124,RN116,RN127,RN56}. OCTExplorer is a prominent example for retinal layer boundary extraction, which is based on a conventional graph-theory algorithm \cite{RN130}. Alonso-Caneiro \emph{et al}. and Tian \emph{et al}. designed graph-theory-based boundary detectors to extract choroidal boundary after image pre-processing \cite{RN129,RN124}. Mayer \emph{et al}. proposed an energy-minimization-based algorithm for retinal nerve fiber layer surface segmentation in circular OCT images \cite{RN123}. Lang \emph{et al}. presented a random forest boundary classifier to segment eight retinal layers in macular cube images \cite{RN121}. Chiu \emph{et al}. reported a kernel regression-based segmentation method for retinal OCT images with diabetic macular edema \cite{RN120}. 

\begin{figure}[h!]
\centering\includegraphics[width=0.95\textwidth]{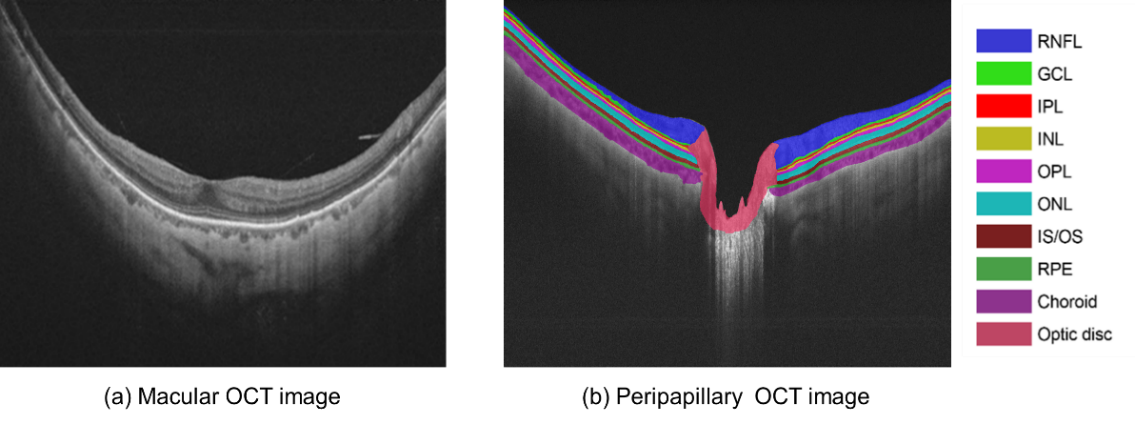}
\caption{The comparison between (a) macular OCT image and (b) peripapillary OCT image. The peripapillary image is manually segmented. Ten labels including RNFL, GCL, IPL, INL, OPL, ONL, IS/OS, RPE, choroid, and optic disc are used and annotated by different colors. The layer structure follows an arrangement that the optic nerve head is located in the center of the image, while much thinner retinal layers are stratified on both sides.}
\end{figure}

Recently, convolutional neural networks (CNN) have been widely applied to segment images obtained from various modalities and thus enable exciting applications \cite{RN148,RN29,RN136,RN149,RN137,RN141,RN152,RN139,RN147}. Fully convolutional networks (FCN) \cite{RN133} and U-Net \cite{RN134} are two popular candidates for medical image segmentation. For retinal OCT image segmentation, most state-of-the-art models \cite{RN58,RN16,RN23,RN153,RN59,RN7,RN4,RN92,RN8,RN44} could be considered variants of the encoder-decoder architecture like FCN and U-Net. Roy \emph{et al}. proposed ReLayNet for end-to-end segmentation of macular OCT B-scans into retinal layers and fluid masses \cite{RN7}. This work is among the first deep learning-based methods for automated segmentation of retinal OCT images. Yang \emph{et al}. designed an attention-guided channel-to-pixel convolution network for retinal layer segmentation with choroidal neovascularization \cite{RN8}: a channel-to-pixel block and an ``edge loss function'' were used to segment the retinal layer with blurry boundaries. To address the large morphological variations of the retinal layers, they also employed the attention mechanism. However, these two techniques were mainly targeting macular retinal image segmentation rather than that of peripapillary retinal images. Zang \emph{et al}. developed an automated segmentation method for peripapillary retinal layer segmentation \cite{RN44}. The left and right boundaries of the optic disc were first determined based on the estimated position of Bruch's membrane opening in radially resampled B-scans. The retinal layer boundaries were then segmented by combining a convolutional neural network with a multi-weight graph search algorithm. Devalla \emph{et al}. proposed a dilated-residual U-Net (DRUNET) to facilitate end-to-end segmentation of the individual neural and connective tissues of the optical nerve head \cite{RN16}. However, they only segmented the retina into five layers and did not fully segment the optic disc from its connected tissues. In summary, most aforementioned techniques perform segmentation based on the textural features of the OCT images, while abundant anatomical priors available in the peripapillary retinal OCT images are not utilized. 

 In this manuscript, we report our recent study on explicit exploiting the prior knowledge existed in the peripapillary OCT images. We argue that all peripapillary OCT images obtained by following a strict clinical protocol should share a similar anatomical arrangement: the optic nerve head, which is a large structure, is located in the center region of the image, while much thinner retinal layers are stratified on both sides, as shown in Fig. 1. Inspired by Jamal \emph{et al}.’s work that uses a graph to represent the domain knowledge and the structural relationship of the tissues \cite{RN155}, we designed a novel multi-scale graph convolutional network (GCN)-assisted two-stage network for joint segmentation of retinal layers and optic disc in peripapillary OCT images to fully take advantage of the anatomical priors. To show the efficacy of the proposed framework, experiments were conducted on a collected peripapillary OCT dataset, which consists of a total number of 122 OCT B-scans from 61 patients, and another public dataset \cite{RN157}. The proposed model demonstrated superior performances on both datasets in comparison with the baselines and the state-of-the-arts. In the future, we plan to integrate the proposed segmentation framework into a diagnostic workflow for early-stage glaucoma detection. The dataset and the source codes are now publicly available online at https://github.com/Jiaxuan-Li/MGU-Net.

\section{Method}
In current study, 10 labels including retinal nerve fiber layer (RNFL), ganglion cell layer (GCL), inner plexiform layer (IPL), inner nuclear layer (INL), outer plexiform layer (OPL), outer nuclear layer (ONL), inner/outer photoreceptor segment (IS/OS), retinal pigment epithelium (RPE), choroid, and optic disc are manually annotated on the OCT dataset to facilitate the training procedure as illustrated in Fig. 1(b).
\subsection{Overview of segmentation framework}
The schematic diagram of the proposed segmentation framework is given in Fig. 2. The entire framework consists of three components: the optic disc detection network, the retinal layer segmentation network, and the fusion module. An input OCT image is first processed by the optic disc detection network, through which a mask indicating the location of the optic disc and the corresponding feature map will be obtained. We then apply the mask on the input image to generate a disc-free image, which is later fed to the retinal layer segmentation network without being sliced.  Similarly, a feature map that delineates the nine retinal layers is also obtained and later concatenated with that of the optic disc from the first stage. Finally, a softmax activation function is used to generate the segmented output based on the concatenated feature map in the fusion module. The entire framework is trained in an end-to-end fashion: two loss functions are defined to penalize both the intermediate disc detection and the final segmentation.

\begin{figure}[h!]
\centering\includegraphics[width=13cm]{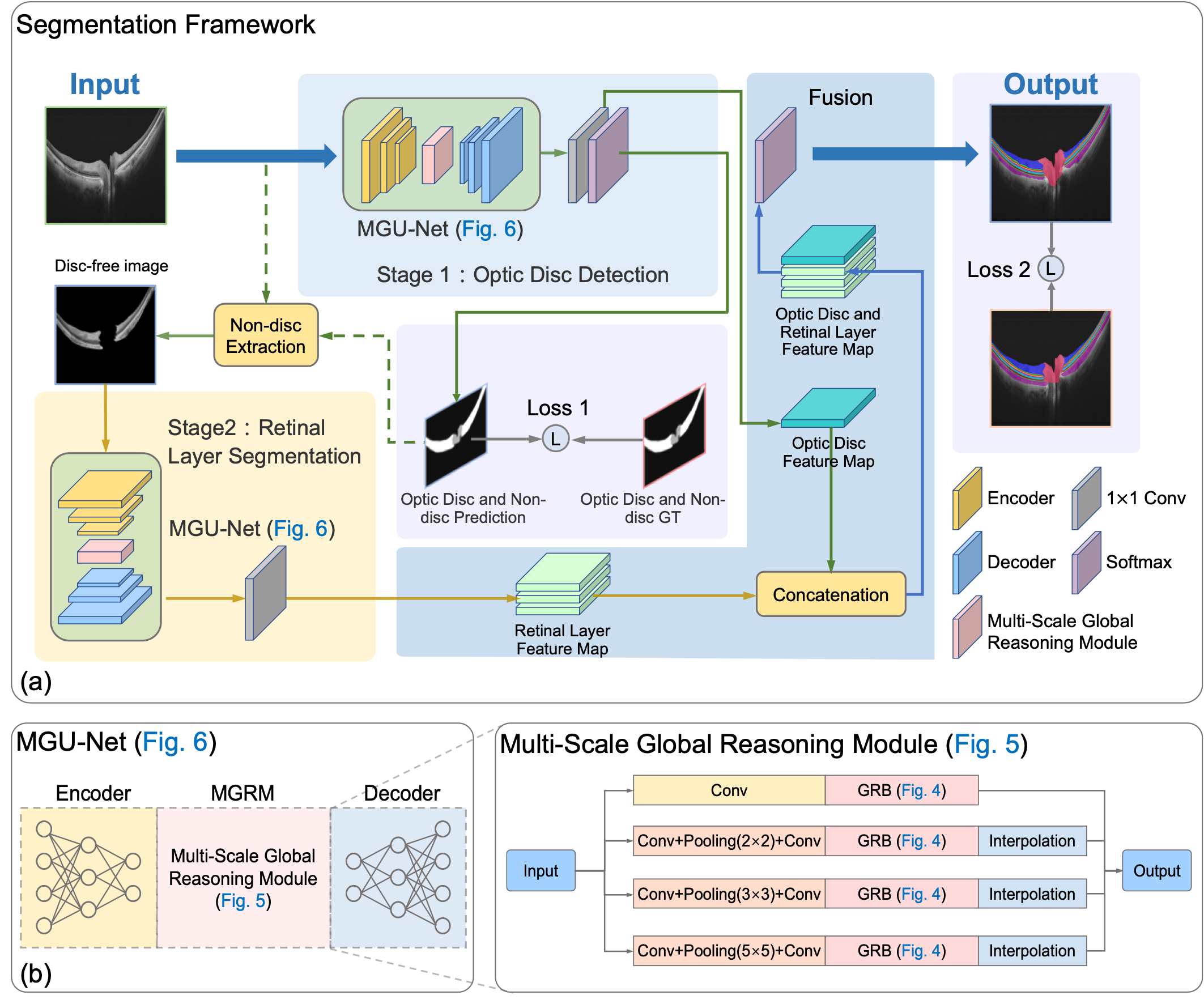}
\caption{Fig. 2. Schematic diagram of the proposed segmentation framework. (a) The entire process consists of three main steps: (1) the first stage is for initial optic disc detection; (2) the second stage is for retinal layer segmentation; (3) the outputs from the previous two stages are later fused. (b) A simplified illustration of Multi-scale GCN-assisted U-shape network (MGU-Net). MGU-Net consists of a pair of encoder and decoder, with a Multi-scale global reasoning module (MGRM) inserted in between. The detailed structure for MGU-Net and MGRM could be found in Fig. 6 and Fig. 5, respectively.}
\end{figure}

Specifically, the optic disc detection network and the retinal layer segmentation network are designed to exploit the anatomical priors of the peripapillary region of the retina and to address the segmentation challenges imposed by the variation of the thicknesses among different retinal layers as illustrated in Fig. 3. In the right panel of Fig. 3, it is observed that the non-disc and disc regions are horizontally arranged in a ``non-disc''-``disc''-``non-disc'' fashion on the global scale. On the other hand, if we zoom in to the ``non-disc'' region as shown in the left panel of Fig. 3, we could appreciate that the retinal layers with various thicknesses are instead vertically arranged. Therefore, the optic disc detection and the retinal layer segmentation networks are devised with graph reasoning blocks with different design goals: the former is to perform long-range horizontal spatial reasoning, while the latter is to capture the multi-level vertical structures. 

\begin{figure}[h!]
\centering\includegraphics[width=\textwidth]{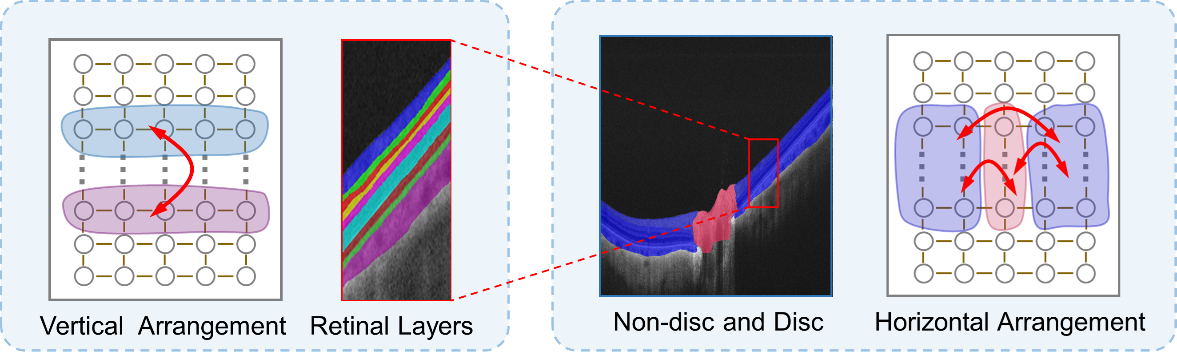}
\caption{ Graph-based representation of peripapillary retinal OCT image. The image possesses a horizontal layout as ``non-disc''-``disc''-``non-disc'' on the global scale. A zoomed-in view of the ``non-disc'' region presents a stratified structure instead. Our segmentation framework is designed to exploit the anatomical priors of the peripapillary region of the retina and to address the segmentation challenges caused by the variation of thickness among retinal layers.}
\end{figure}

\subsection{Graph reasoning block}
The key module used in both the optic disc detection network and the retinal layer segmentation network is the graph reasoning block.

\begin{figure}[h!]
\centering\includegraphics[width=0.95\textwidth]{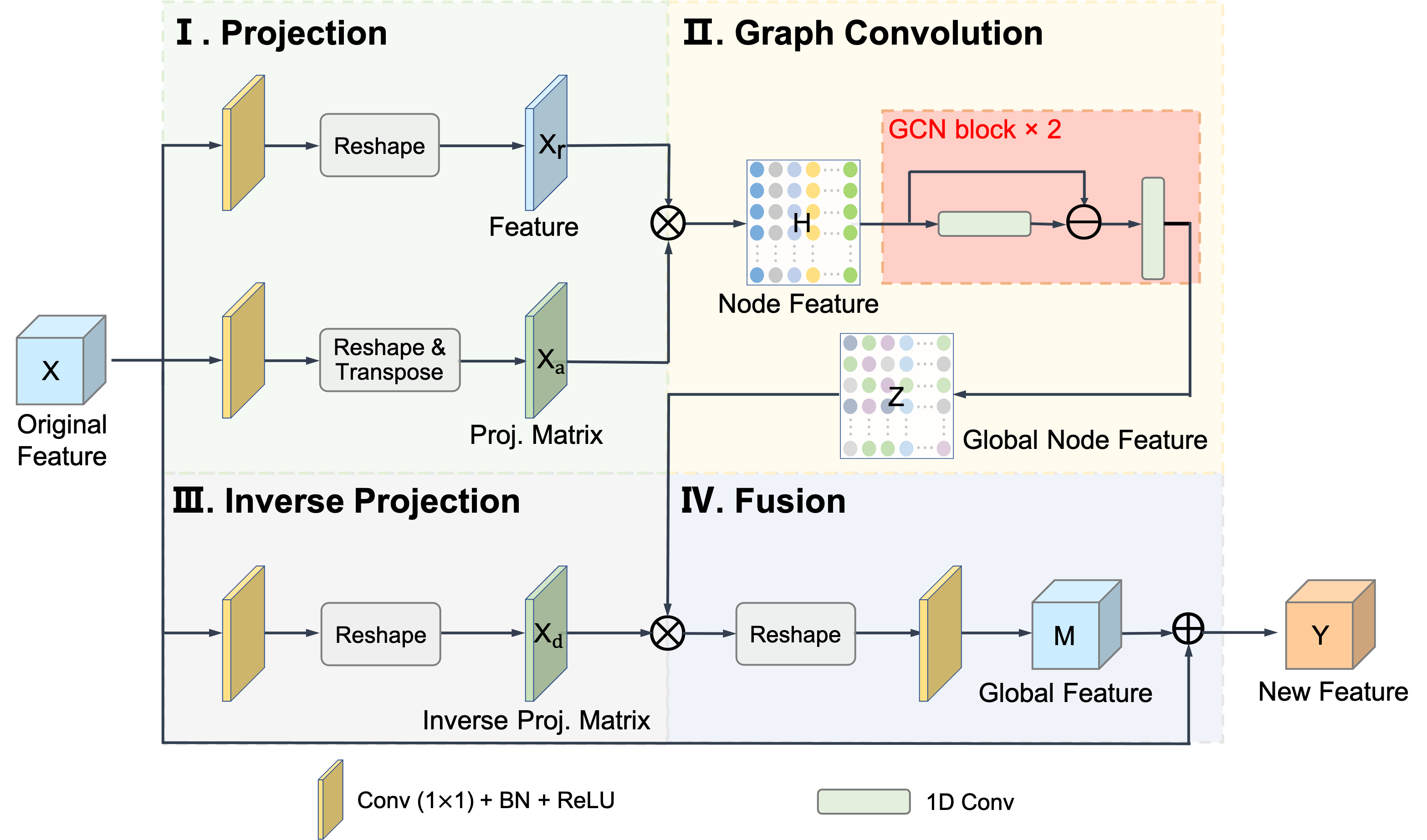}
\caption{The schematic diagram of a graph reasoning block, which consists of four operations. First of all, the original features are projected to the node space. After that, graph convolutions are performed on the node-space features to extract global node features. In order to fuse the global node features with the original features, they are then inversely projected back to the original feature space before being fused with the original features.}
\end{figure}

Inspired by the graph-based global reasoning network \cite{RN5,RN60}, we devised a graph reasoning block to effectively extract the global features of 9 retinal layers and optic disc. The schematic diagram of the graph reasoning block, which consists of four operations, is depicted in Fig. 4. First of all, a local feature map $\mathrm{X} \in \mathbb{R}^{\mathrm{C} \times \mathrm{H} \times \mathrm{W}}$ in the latent space is fed to two convolutional layers in parallel to generate two maps: one feature map with reduced dimension and one projection matrix. After that, the reduced dimension feature is reshaped to $\mathrm{X}_{\mathrm{r}} \in \mathbb{R}^{\mathrm{C}_{\mathrm{r}} \times \mathrm{HW}},$ while the projection matrix is reshaped and transposed as $\mathrm{X}_{\mathrm{a}} \in \mathbb{R}^{\mathrm{C}_{\mathrm{n}} \times \mathrm{HW}}$. A matrix multiplication between $\mathrm{X}_{\mathrm{r}}$ and $\mathrm{X}_{\mathrm{a}}$ is then performed to obtain a node feature map $\mathrm{H} \in \mathbb{R}^{\mathrm{C}_{\mathrm{r}} \times \mathrm{C}_{\mathrm{n}}}$ before its being sent to a GCN block. We further connect $\mathrm{X}$ to a convolutional layer to create an inverse projection matrix $\mathrm{X}_{\mathrm{d}} \in \mathbb{R}^{\mathrm{C}_{\mathrm{n}} \times \mathrm{H} \times \mathrm{W}}$. The output of the GCN block is multiplied by the reshaped $\mathrm{X}_{\mathrm{d}} \in \mathbb{R}^{\mathrm{C}_{\mathrm{n}} \times \mathrm{H} \times \mathrm{W}}$ to transform back to the original latent space, which is then reshaped and undergo another convolutional layer to eventually obtained the feature map $\mathrm{M} \in \mathbb{R}^{\mathrm{C} \times \mathrm{H} \times \mathrm{W}}$. Finally, we perform an element-wise addition of $\mathrm{M}$ and $\mathrm{X}$ to acquire the new feature map $\mathrm{Y} \in \mathbb{R}^{\mathrm{C} \times \mathrm{H} \times \mathrm{W}}$

It is clear that the new feature map Y contains both the information from the global feature and the original feature, which enables its capability of processing long-range contextual information.

\subsection{Multi-scale global reasoning module}
To address the segmentation challenges caused by the large variation of the retinal thicknesses between different layers, we propose a multi-scale global reasoning module (MGRM) to conduct global reasoning on high-level semantic features in all nine retinal layers. MGRM is composed of multi-scale pooling operators and graph reasoning blocks. It uses multiple effective receptive fields to capture and learn the features of the retinal layers with different thicknesses.

The MGRM split the input into four different paths, three of which are equipped with pooling layers of different kernel sizes followed by graph reasoning blocks: inspired by \cite{RN28}, the kernel sizes are set to  2 × 2, 3 × 3 and 5 × 5 to encode the information of retinal layers with different thicknesses to global feature maps. Then, the global features are up-sampled to match the size of the original input feature map by bilinear interpolation. Finally, all four paths are re-combined and the features are concatenated. The entire procedure is illustrated in Fig. 5.

\begin{figure}[h!]
\centering\includegraphics[width=0.8\textwidth]{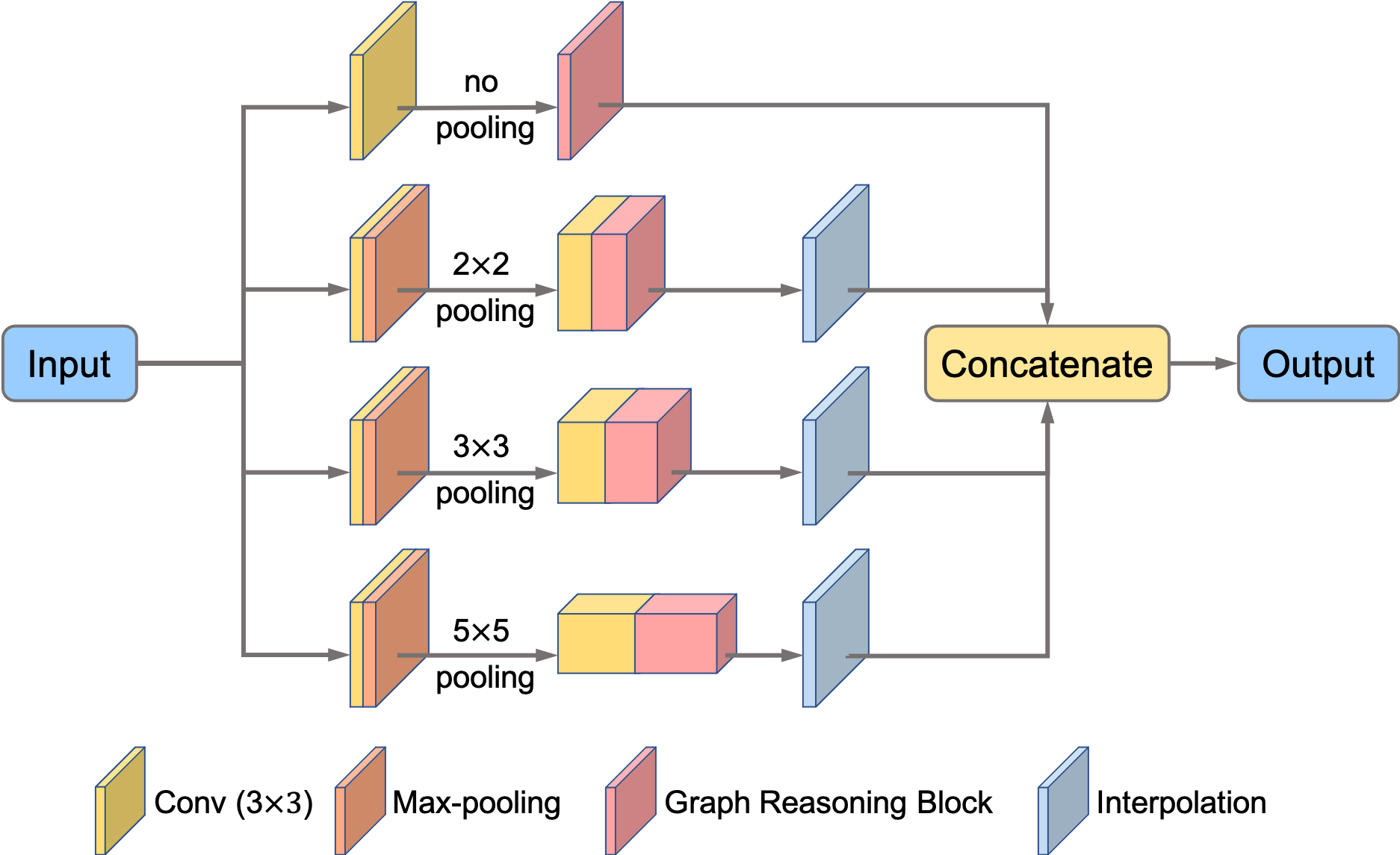}
\caption{The structure of multi-scale global reasoning module is composed of four branches. No pooling operator is in the first branch. There are pooling operators with 2 × 2 kernel, 3 × 3 kernel and 5 × 5 kernel in the second, third and fourth branch.}
\end{figure}

\subsection{Multi-scale GCN-assisted U-shape network}
We use a U-shape network developed on the basis of the classic U-Net \cite{RN134} as the backbone of the proposed multi-scale GCN-assisted U-shape network (MGU-Net). The schematic diagram of MGU-Net is presented in Fig. 6. MGRM is located in the center of the network to connect the encoding and decoding paths. It captures additional long-range contextual features, which are difficult to acquire in conventional neural network. After several convolution operations and max-pooling operations in the encoder part, the feature map provides rich spatial features, which are informative for aggregating features and extracting nodes in the following MGRM. It should be noted that the sizes of the max-pooling kernels are different for the optic disc detection network and the retinal layer segmentation network: the size of each max-pooling kernel is set to (2, 2, 2) in the retinal layer segmentation network as in Fig. 6, while that of the optic disc detection network is set to be (2, 4, 2) to better capture the larger-scale semantic information represented by the optic disc.

\begin{figure}[h!]
\centering\includegraphics[width=0.8\textwidth]{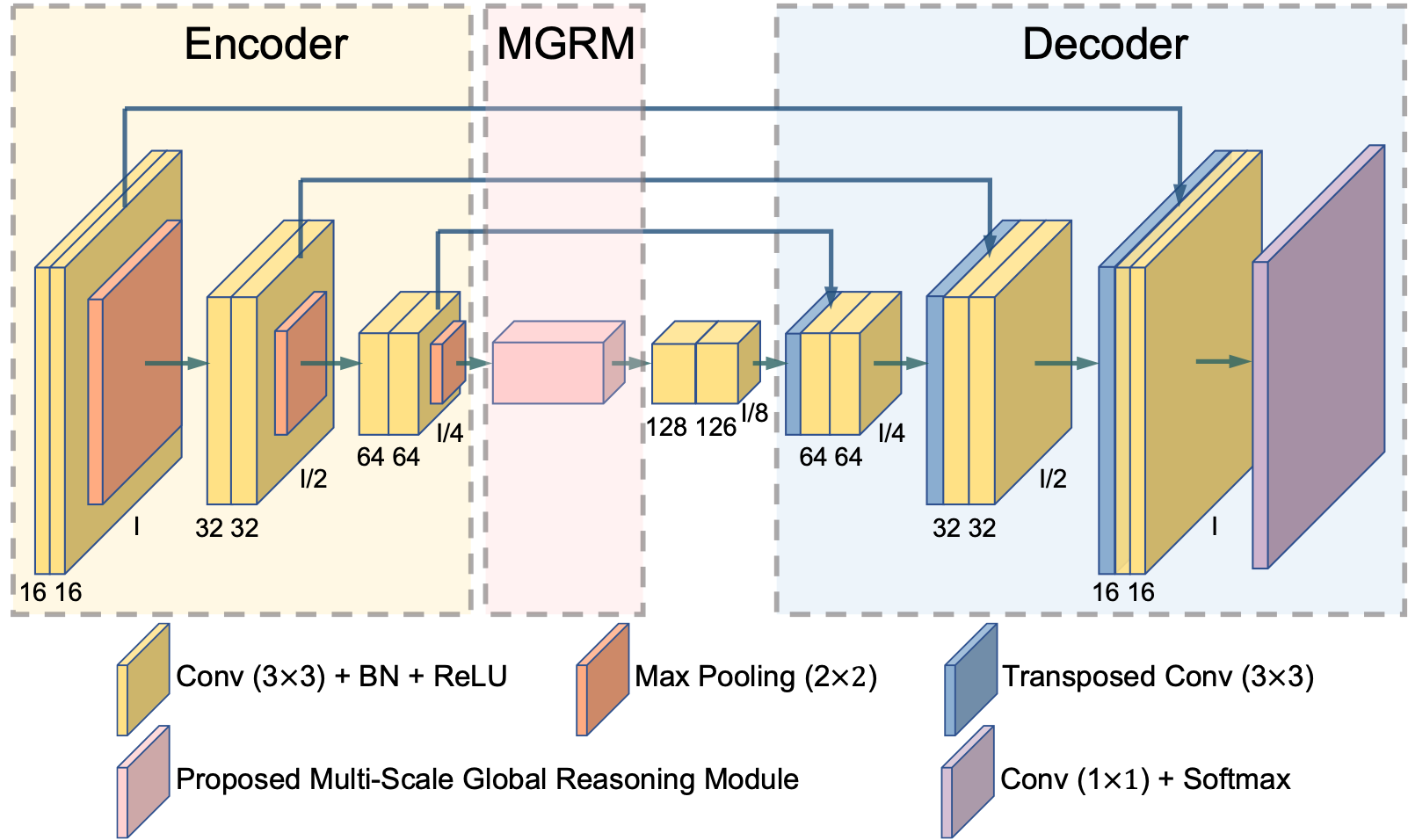}
\caption{The structure of MGU-Net comprises of encoder, MGRM and decoder. The skip-connections concatenate low-level features from encoding path to the corresponding high-level features in decoding path.}
\end{figure}

\subsection{Loss function}
In our two-stage segmentation framework, two loss functions $\mathcal{L}_{\mathrm{Seg}_{1}}$ and $\mathcal{L}_{\mathrm{seg}_{2}}$ are proposed to
supervise these two stage MGU-Nets and to enforce them to segment optic disc and nine retinal layers in an end-to-end fashion more accurately. The total loss $\mathcal{L}$ in this study is the sum of the two losses. The total loss function is shown as follow,
\begin{equation}
\mathcal{L}= \mathcal{L}_{\mathrm{seg}_{1}}+\lambda\mathcal{L}_{\mathrm{seg}_{2}}
\end{equation}
where $\lambda$ weights the two losses.
$\mathcal{L}_{\text {seg are }}$ defined as the sum of Dice loss and Cross-Entropy loss, which can be described as
\begin{equation}
\mathcal{L}_{\text {seg }}=\mathcal{L}_{\text {dioe }}+\mathcal{L}_{\text {ce }}
\end{equation}
where
\begin{equation}
\mathcal{L}_{\text {dice }}=1-\frac{1}{M} \sum_{i=1}^{M}\left(\frac{2 \sum_{x \in \Omega} p_{i}(x) \times g_{i}(x)}{\sum_{x \in \Omega} p_{i}(x)+\sum_{x \in \Omega} g_{i}(x)}\right) \\
\mathcal{L}_{\mathrm{ce}}=-\frac{1}{M} \sum_{i=1}^{M} g_{i} \log \left(p_{i}\right)
\end{equation}
in which $g_{i}$ and $p_{i}$ indicates the ground truth and the probability in prediction of pixel $x$ belonging
to of class $i, M$ is the number of classes in the segmentation network.

\section{Experiment Design}
\subsection{Datasets and implementation details}
\subsubsection{Collected dataset}
To verify the effectiveness of the framework, we conducted a series of experiments on collected peripapillary retinal OCT images. All images were de-identified and the procedure was approved by Internal Review Board of Shanghai General Hospital. The entire dataset consists of 61 different subjects, for each of which 12 radial OCT B-scans are collected at the Ophthalmology Department of Shanghai General Hospital by using DRI OCT-1 Atlantis (Topcon Corporation, Tokyo, Japan). The clinical characteristics of dataset are provided in Table 1. The image size is 1024 × 992 in pixel, corresponding to a field of view of 20.48 mm × 7.94 mm.

For each subject, 2 radial OCT B-scans were randomly selected to ensure mutual exclusion. Two graders annotated these images manually through ITK-SNAP software \cite{RN135} into optic disc and nine retinal layers under the supervision of a glaucoma specialist. While one image is only annotated by one grader, the final ground truth is obtained from the consensus of all personnel.

For the experiment, the data were randomly divided into three mutually exclusive subsets for training, validation, and testing on the patient level. The ratio between these sets was 6:2:2. In order to enlarge the size of our dataset, we performed data augmentation including horizontal flipping, additive Gaussian noises, and  contrast adjustment.

\begin{table}[h!]
\caption{Clinical characteristics of subjects in dataset.} 
\small 
\centering
\begin{tabular}{lc}
\hline
\textbf{Characteristics}                              & \textbf{Value} \\ \hline
Subject                                               & 61             \\ \hline
\multicolumn{2}{l}{\textbf{Demographic profile}}                       \\
Male, n (\%)                                          & 26 (42.6\%)    \\
Age in yrs,   mean ± standard deviation               & 66.40 ± 6.91   \\
Ocular   pressure in mm Hg, mean ± standard deviation & 12.52 ± 3.23   \\
Axis length   in mm, mean ± standard deviation        & 26.53 ± 1.66   \\ \hline
\multicolumn{2}{l}{\textbf{Ophthalmological disease}}                  \\
High myopia,   n (\%)                                 & 34 (55.7\%)    \\
Peripapillary   atrophy, n (\%)                       & 38 (62.3\%)    \\
Cataract, n   (\%)                                    & 29 (47.5\%)    \\ \hline
\end{tabular}
\end{table}

\subsubsection{Public dataset}
In addition, we tested our proposed technique on the Duke SD-OCT dataset, which was collected by Chiu \emph{et al}. using a Spectralis HRA+OCT (Heidelberg Engineering, Heidelberg, Germany) \cite{RN157}. It consists of 110 OCT B-scans obtained from 10 patients with diabetic macular edema (DME) with a size of 496 × 768 pixels. More details about this dataset could be found in \cite{RN157}.

\subsubsection{Experimental details}
The proposed method was implemented in PyTorch and trained on NVIDIA Tesla V100 GPUs. During the training, the initial learning rate was 0.001 and was reduced by an order of magnitude after every 20 epochs. The number of epochs was 50. Momentum and weight decay coefficients were set to 0.9 and 0.0001, respectively. We used Adam optimizer to train the model in mini batches of size = 1. Parameter $\lambda$ in Eq. (1) is empirically set as 2. To ensure a fair comparison, the training hyperparameters were kept constant to achieve the best performance for all the comparative methods.

\subsection{Comparisons with the state-of-the-arts on collected dataset}
We compared our model with state-of-the-art techniques, including U-Net \cite{RN134}, ReLayNet \cite{RN7} and DRUNET \cite{RN16}. U-Net is a popular segmentation network used for medical image. ReLayNet is specially designed for segmenting retinal layers and fluid in macular OCT images. DRUNET is proposed to segment optic nerve head tissues in peripapillary OCT images using a dilated and residual U-shape network. It is worth noting that U-Net has four down-sampling and four up-sampling operators, while ReLayNet and DRUNET both perform three down-sampling and three up-sampling operations.

\subsection{Comparisons with state-of-the-arts on public dataset}
We repeat the experiments on the Duke SD-OCT dataset, which is macular centered. Since the proposed two-stage framework is originally designed for segmenting peripapillary OCT B-scans, we removed the first stage and only use the second stage to compete against the state-of-the-art models including U-Net, ReLayNet, DRUNET and published results on this public dataset  \cite{RN158,RN157,RN7,RN4}.

\subsection{Ablation study}
To assess the contribution of each component of the proposed framework, we performed several ablation studies on collected dataset. We compare the  performance of the proposed model with that of (1) one-stage baseline, (2) two-stage baseline, (3) without graph reasoning blocks in multi-scale global reasoning module, and (4) with single-scale global reasoning module in the two-stage segmentation framework.

\subsection{Evaluation Metrics}
The Dice score (DSC) and pixel accuracy (PA) between predictions and segmentation references were used for quantitative evaluation of segmentation performance. They are calculated as:

\begin{equation}
\operatorname{DSC} =\frac{2|X \cap Y|}{|X|+|Y|} 
\end{equation}
and
\begin{equation}
\mathrm{PA} =\frac{|X \cap Y|}{|Y|}
\end{equation}
where $X$ is the region of prediction and $Y$ is the region of ground truth. Dice score  was used to measure the overlap between the prediction and ground truth. Pixel accuracy was used to calculate the true positive rate of predicted results compared with their ground truth.

\section{Results}
\subsection{Collected dataset}
The experimental results on the collected dataset are listed in Table 2 and Table 3. The proposed method outperforms the selected state-of-the-art methods in most optic nerve head tissue categories except for the optic disc (both Dice and pixel accuracy) and RPE (pixel accuracy): the overall average Dice score for the proposed network is 1.6$\%$, 1.5$\%$, and 1.4$\%$ higher than that of ReLayNet, U-Net, and DRUNET, respectively. A similar trend is observed in the pixel accuracy results.

\begin{table}[h!]
\centering
\caption{Dice score ($\%$) for the segmentation on collected dataset obtained by different methods. The best performance is marked by “*”, the second-best performance is indicated by “**”. Improvement is defined as the difference between the proposed method and the best performance obtained among other techniques.} 
\small 
\begin{tabular}{lccccclll}
\cline{1-6}
\textbf{}                 & \multicolumn{4}{c}{\textbf{Method}}                                      & \textbf{}            &  &  &  \\ \cline{2-5}
\textbf{Tissue}                    & \textbf{U-Net} & \textbf{ReLayNet} & \textbf{DRUNET} & \textbf{Proposed} & \textbf{Improvement} &  &  &  \\ \cline{1-6}
\textbf{Average}          & 80.5±0.4       & 80.4±0.4          & 80.6±0.5**      & 82.0±0.1*         & ↑ 1.4                &  &  &  \\ \cline{1-6}
\textbf{Layer}            & 80.2±0.4       & 80.9±0.4**        & 80.9±0.5        & 82.0±0.2*         & ↑ 1.1                &  &  &  \\ \cline{1-6}
\enskip  \emph{RNFL}    & 82.0±0.6       & 81.6±0.3          & 82.4±0.3**      & 83.1±0.2*         & ↑ 0.7                &  &  &  \\ \cline{1-6}
\enskip  \emph{GCL}     & 66.8±1.3       & 68.6±0.7**        & 68.3±0.7        & 69.9±0.3*         & ↑ 1.3                &  &  &  \\ \cline{1-6}
\enskip  \emph{IPL}    & 72.0±0.8       & 73.0±1.2**        & 72.7±1.1        & 74.9±0.5*         & ↑ 1.9                &  &  &  \\ \cline{1-6}
\enskip  \emph{INL}    & 75.1±0.8       & 77.3±1.2**        & 76.9±1.0        & 78.9±0.6*         & ↑ 1.6                &  &  &  \\ \cline{1-6}
\enskip  \emph{OPL}     & 79.2±0.7       & 80.8±0.7**        & 80.5±0.9        & 82.0±0.4*         & ↑ 1.5                &  &  &  \\ \cline{1-6}
\enskip \emph{ONL}    & 90.5±0.1       & 90.4±0.3          & 90.6±0.2**      & 90.9±0.1*         & ↑ 0.3                &  &  &  \\ \cline{1-6}
\enskip \emph{IS/OS}  & 85.7±0.2       & 86.2±0.2*         & 85.8±0.3**      & 86.2±0.2*         & ↑ 0.0                &  &  &  \\ \cline{1-6}
\enskip \emph{RPE}    & 82.2±0.2       & 82.4±0.5**        & 82.2±0.3        & 82.6±0.1*         & ↑ 0.2                &  &  &  \\ \cline{1-6}
\enskip \emph{Choroid} & 88.3±0.5**     & 87.7±0.2          & 88.2±0.5        & 89.2±0.2*         & ↑ 0.9                &  &  &  \\ \cline{1-6}
\textbf{Disc}             & 83.0±0.7*      & 75.9±0.6          & 78.5±1.9        & 82.1±0.7          & ↓ 0.9                &  &  &  \\ \cline{1-6}
\end{tabular}
\end{table}

\begin{table}[h!]
\centering
\caption{Pixel accuracy ($\%$) of segmentation results on collected dataset by different methods and improvement in comparison with the best performance from other methods. The best performance is marked by “*”, the second-best performance is indicated by “**”.} 
\small 
\begin{tabular}{lccccclll}
\cline{1-6}
\textbf{}                 & \multicolumn{4}{c}{\textbf{Method}}                                      &                      &  &  &  \\ \cline{2-5}
\textbf{Tissue}           & \textbf{U-Net} & \textbf{ReLayNet} & \textbf{DRUNET} & \textbf{Proposed} & \textbf{Improvement} &  &  &  \\ \cline{1-6}
\textbf{Average}          & 81.2±0.6**     & 80.2±0.5          & 81.1±0.4        & 83.0±0.2*         & ↑ 1.8                &  &  &  \\ \cline{1-6}
\textbf{Layer}            & 80.6±0.6**     & 80.5±0.8          & 80.6±0.7        & 82.6±0.3*         & ↑ 2.0                &  &  &  \\ \cline{1-6}
\enskip \emph{RNFL}    & 83.1±0.9**     & 81.8±0.3          & 82.1±0.6        & 84.6±1.6*         & ↑ 1.5                &  &  &  \\ \cline{1-6}
\enskip \emph{GCL}     & 66.9±2.8       & 68.7±1.7**        & 67.9±0.8        & 70.3±1.8*         & ↑ 1.6                &  &  &  \\ \cline{1-6}
\enskip \emph{IPL}   & 72.1±1.5**     & 70.9±1.7          & 71.4±1.3        & 74.7±1.5*         & ↑ 2.6                &  &  &  \\ \cline{1-6}
\enskip \emph{INL}    & 76.0±0.8       & 77.9±2.0**        & 77.8±1.2        & 81.0±1.9*         & ↑ 3.1                &  &  &  \\ \cline{1-6}
\enskip \emph{OPL}    & 78.3±1.7       & 79.7±1.5**        & 79.5±1.6        & 82.0±1.0*         & ↑ 2.3                &  &  &  \\ \cline{1-6}
\enskip \emph{ONL}     & 91.4±0.6       & 90.3±1.2          & 91.9±0.5**      & 91.9±0.5*         & ↑ 0.0                &  &  &  \\ \cline{1-6}
\enskip \emph{IS/OS}   & 86.2±0.7       & 86.6±0.8**        & 85.7±0.6        & 87.0±0.4*         & ↑ 0.4                &  &  &  \\ \cline{1-6}
\enskip \emph{RPE}     & 83.1±0.6*      & 82.3±2.4          & 81.9±1.2        & 83.0±0.4**        & ↓ 0.1                &  &  &  \\ \cline{1-6}
\enskip \emph{Choroid} & 88.2±1.4**     & 86.6±0.8          & 87.4±1.9        & 89.2±0.7*         & ↑ 1.0                &  &  &  \\ \cline{1-6}
\textbf{Disc}             & 86.8±1.5*      & 77.5±2.2          & 85.6±3.3        & 86.1±1.2**        & ↓ 0.7                &  &  &  \\ \cline{1-6}
\end{tabular}
\end{table}

\begin{figure}[h!]
\centering\includegraphics[width=0.95\textwidth]{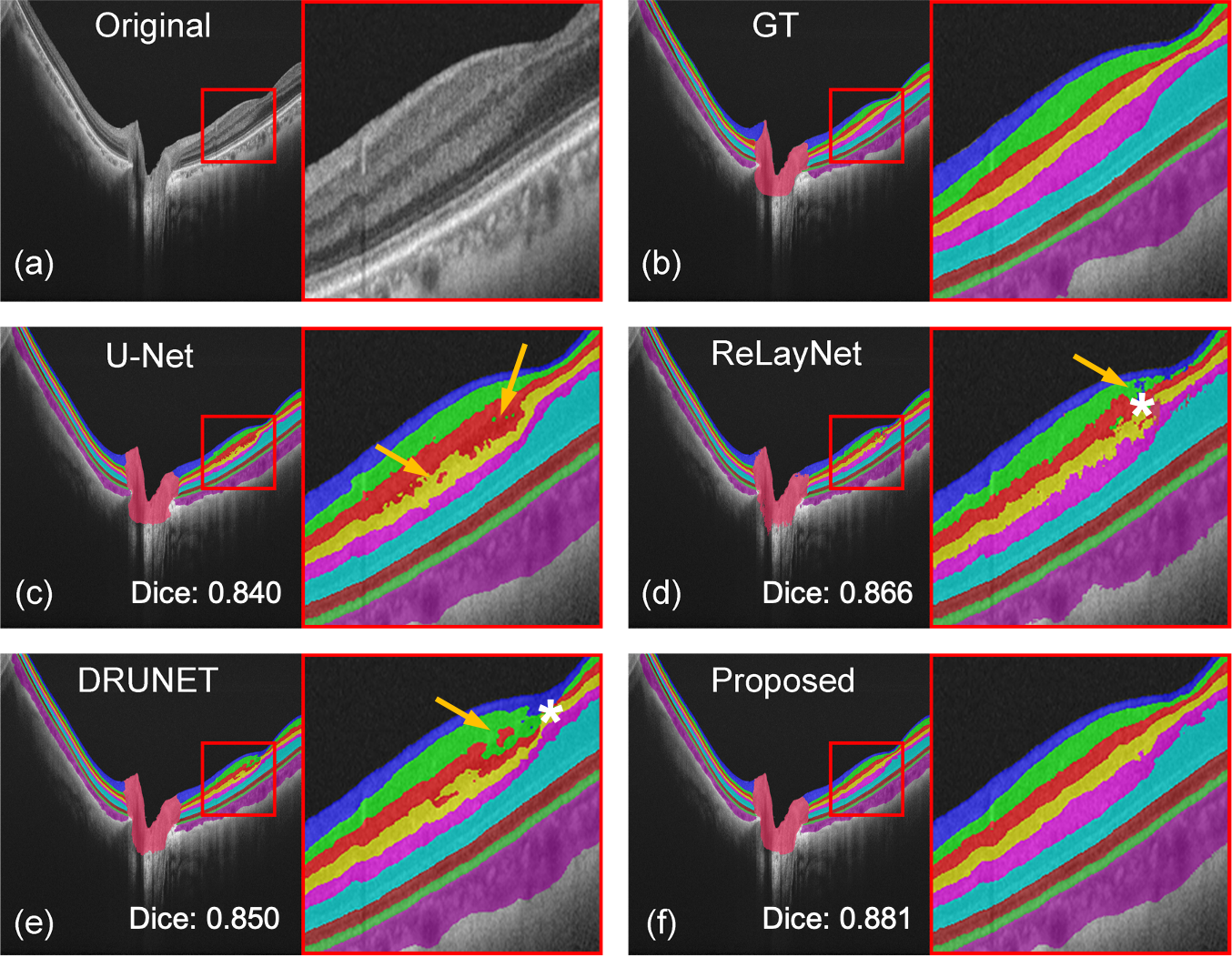}
\caption{Segmentation of a normal peripapillary OCT image. (a) Original image. (b) Ground truth. (c) U-Net’s prediction. (d) ReLayNet’s prediction. (e) DRUNET’s prediction. (f) Proposed method’s prediction. Exemplary scattered labels are pointed out by the yellow arrows, while layer discontinuities are marked by white stars. Magnified views are provided for better visualization.}
\end{figure}

Fig. 7 shows the segmentation results obtained by various techniques on a normal peripapillary OCT image. The segmented image obtained by the proposed method as shown in Fig.7(f) presents the best visual quality, while various artifacts are visible in the others. We roughly categorize the artifacts into two groups,

1)	layer discontinuities, where the stratified retinal layers are unexpectedly terminated in the horizontal direction as observed in ReLayNet (Fig. 7(d)) and DRUNET (Fig. 7(e)) and marked by white stars;

2)	scattered labels, where an isolated region enclosed by a certain layer to be recognized as others as pointed out by the yellow arrows in Fig. 7(c)-(d). This type of errors is observed in all but the proposed method.

We suggest that the improved performance could be ascribed to the additional prior knowledge incorporated in the proposed framework. For conventional CNN-based algorithms, the pixel-level classification is often sensitive to the textural details, while we regularize this with extra spatial constraints in the proposed technique: the segmentation results have to comply with the learned spatial layout, which dictates that the retinal layers must be arranged as a horizontally continuous and vertically stratified structure.

A similar observation could be made on a diseased sample with retinal lesion as illustrated in Fig. 8. It is clear that the blurred boundaries and the reduced contrast in the lesion area as circled out in Fig. 8(a) are challenging for conventional CNN-based algorithms. On the other hand, while mis-labelling are also observed in the proposed method, the stratified structure of the retinal layers is well preserved, which again demonstrates the efficacy of the proposed technique.

\begin{figure}[h!]
\centering\includegraphics[width=0.95\textwidth]{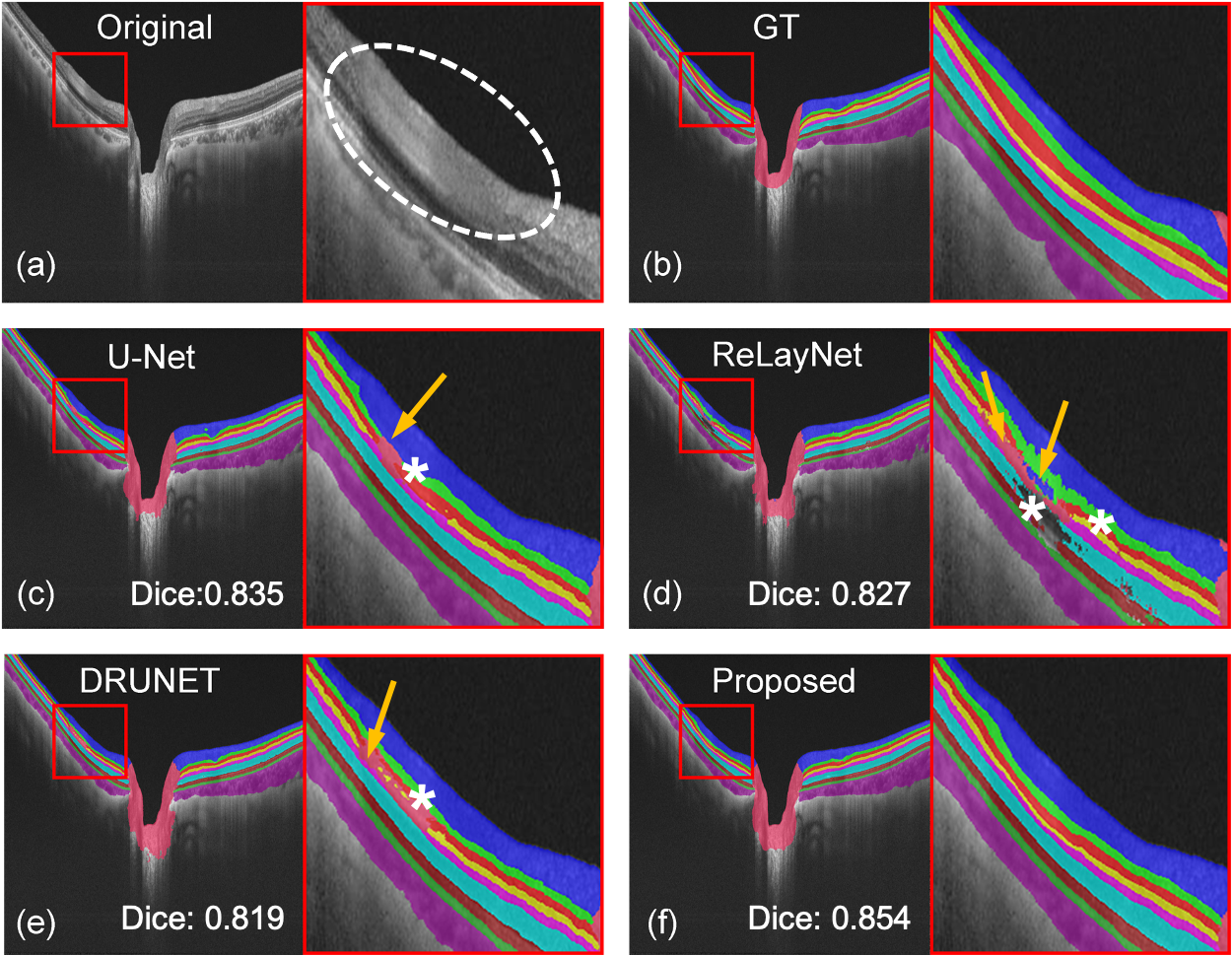}
\caption{Segmentation of a peripapillary OCT image with retinal lesion. (a) Original image. (b) Ground truth. (c) U-Net’s prediction. (d) ReLayNet’s prediction. (e) DRUNET’s prediction. (f) Proposed method’s prediction. The scattered labels are pointed out by yellow arrows, the layer discontinuities are marked by white star, and the retinal lesion is circled out by white dashed line. Magnified views are also provided for better visualization.}
\end{figure}

\subsection{Public dataset}
The experimental results obtained on the Duke SD-OCT dataset are reported in Table 4. It is worth mentioning that the Duke dataset is macular-centered, while our segmentation framework is designed for disc-center images. Therefore, only one stage of proposed MGU-Net is used in this experiment. Nonetheless, the proposed model achieved best performance in ONL-ISM layer and second-best performance in four retinal layers. The average Dice score achieved by the proposed MGU-Net is the highest if we do not take the results reported by Roy et al into account, while it does outperform the ReLayNet we reproduced. 
We also displayed the segmented images in Fig. 9. The proposed MGU-Net manifests better visual quality in comparison with other OCT retinal image segmentation methods. Consistent with the observations we have made in Section 4.1, artifacts such as layer discontinuities, which are marked by white stars, are presented in the image segmented by U-Net and DRUNET in Fig. 9(c) and Fig. 9(e), respectively.

\begin{table}[h!]
\centering
\caption{Dice score for segmentation results on Duke SD-OCT dataset by different methods and expert 2 annotations. The best performance is marked by “*”, the second-best performance is indicated by “**”.} 
\small 
\begin{tabular}{lcccccccc}
\hline
                         & \multicolumn{7}{c}{\textbf{Tissue}}                                                                                & \textbf{}        \\ \cline{2-8}
\textbf{Method}          & \textbf{RNFL} & \textbf{GCL-IPL} & \textbf{INL} & \textbf{OPL} & \textbf{ONL-ISM} & \textbf{ISE} & \textbf{OS-RPE} & \textbf{Average} \\ \hline
\textbf{Manual expert 2}\cite{RN158} & 0.86          & 0.89             & 0.8          & 0.72         & 0.88             & 0.86         & 0.84            & 0.84             \\ \hline
\textbf{Chiu}\cite{RN157}      & 0.86          & 0.88             & 0.73         & 0.73         & 0.86             & 0.86         & 0.8             & 0.82             \\ \hline
\textbf{Chakravarty}\cite{RN158}  & 0.86          & 0.89             & 0.8          & 0.72         & 0.88             & 0.86         & 0.84            & 0.84             \\ \hline
\textbf{U-Net}\cite{RN7}     & 0.86          & 0.91             & 0.83**       & 0.81**       & 0.91             & 0.9          & 0.83            & 0.86             \\ \hline
\textbf{Roy}\cite{RN7}    & 0.90*         & 0.94*            & 0.87*        & 0.84*        & 0.93             & 0.92*        & 0.90*           & 0.90*            \\ \hline
\textbf{Wang}\cite{RN4}     & 0.86          & 0.9              & 0.78         & 0.78         & 0.94**           & 0.9          & 0.86            & 0.86             \\ \hline
\textbf{U-Net}           & 0.87          & 0.91             & 0.797        & 0.779        & 0.937            & 0.897        & 0.863           & 0.865            \\ \hline
\textbf{ReLayNet}        & 0.872         & 0.905            & 0.807        & 0.782        & 0.939            & 0.898        & 0.862           & 0.867            \\ \hline
\textbf{DRUNET}          & 0.851         & 0.881            & 0.72         & 0.726        & 0.919            & 0.887        & 0.847           & 0.833            \\ \hline
\textbf{MGU-Net (ours)}  & 0.874**       & 0.911**          & 0.81         & 0.792        & 0.943*           & 0.901**      & 0.865**         & 0.871**          \\ \hline
\end{tabular}
\end{table}

\begin{figure}[h!]
\centering\includegraphics[width=13cm]{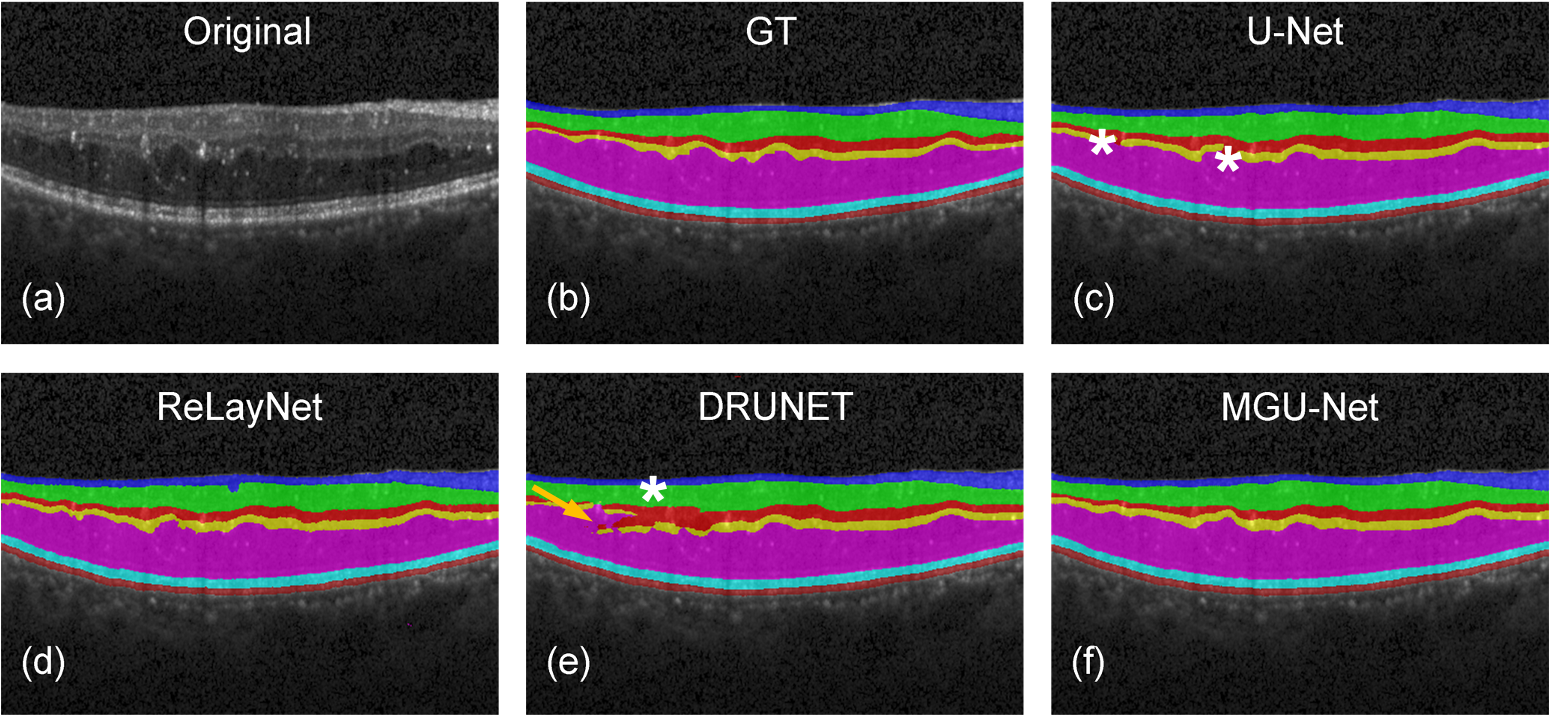}
\caption{Visualization of segmentation results on Duke SD-OCT dataset. (a) Original image. (b) Ground truth. (c) U-Net’s prediction. (d) ReLayNet’s prediction. (e) DRUNET’s prediction. (f) Our proposed method’s prediction. The scattered labels are pointed out by yellow arrow, the discontinuities are marked with white star.}
\end{figure}

\subsection{Ablation study on the collected dataset}
The quantitative results of the ablation study on our dataset are listed in Table 5 and Table 6. Baseline is the U-shape network with three down-sampling operations, which is one level shallower than the U-Net used in previous section. We first compared the results of two-stage baseline with that of one-stage baseline. If a two-stage network is adopted to segment optic disc and retinal layers separately, the Dice score and pixel accuracy were improved by 0.4$\%$ and 1.0$\%$ respectively. If we insert single-scale graph reasoning block (GRB) into the two-stage baseline, the Dice scores were greatly increased compared with two-stage baseline. However, pixel accuracy on RNFL dropped slightly. On the other hand, if we apply multi-scale pooling (MSP) to the two-stage baseline, an immediate boost is also observed from all the results except for the pixel accuracy on RNFL. We suspect that the lack of improvement by adding GRB or MSP might be ascribed to the complicated morphology of the image: single-scale GRB could not capture all spatial information at one time, while MSP could not perform spatial reasoning in global level. After introducing the multi-scale global reasoning module into the two-stage network, which is the combination of GRB and MSP, all the Dice score and pixel accuracy were both lifted, achieved an improvement of 0.9$\%$ and 1.2$\%$ compared with that of two-stage baseline. The results of ablation study demonstrate the effective of the added modules and indicate that the proposed multi-scale GCN-assisted two-stage U-shape network significantly improved the performance of segmentation in peripapillary OCT image.

\begin{table}[h!]
\centering
\caption{Ablation study of each parts in our framework through comparing the Dice score ($\%$). The best performance is marked by “*” ($\%$).} 
\small 
\begin{tabular}{llllllll}
\hline
\textbf{Method}             & \textbf{One-stage} & \textbf{Two-stage} & \textbf{GRB} & \textbf{MSP} & \textbf{AVG.}      & \textbf{RNFL}      & \textbf{GCL}     \\\hline
\textbf{Baseline}           & \checkmark         &           &     &     & 80.7±0.2  & 81.7±0.2  & 68.1±0.9  \\\hline
\textbf{Two-stage baseline} & \checkmark        & \checkmark         &     &     & 81.1±0.3  & 82.2±0.3  & 68.6±1.0  \\\hline
\textbf{Proposed w/o MSP} & \checkmark        & \checkmark         & \checkmark   &     & 81.8±0.1  & 83.0±0.3  & 69.5±0.4  \\\hline
\textbf{Proposed w/o GRB} & \checkmark         & \checkmark         &     & \checkmark   & 81.6±0.3  & 82.7±0.1  & 69.6±0.5  \\\hline
\textbf{Proposed}          & \checkmark         & \checkmark         & \checkmark   & \checkmark   & 82.0±0.1* & 83.1±0.2* & 69.9±0.3* \\\hline
\end{tabular}
\end{table}

\begin{table}[h!]
\centering
\caption{Ablation study of each parts in our framework through comparing the pixel accuracy ($\%$). The best performance is marked by “*” ($\%$).} 
\small 
\begin{tabular}{llllllll}
\hline
\textbf{Method}             & \textbf{One-stage} & \textbf{Two-stage} & \textbf{GRB} & \textbf{MSP} & \textbf{AVG.}      & \textbf{RNFL}      & \textbf{GCL}     \\\hline
\textbf{Baseline}   & \checkmark         &           &     &     & 80.8±0.3  & 81.9±1.8  & 67.1±2.3  \\\hline
\textbf{Two-stage baseline} & \checkmark         & \checkmark         &     &     & 81.8±0.5  & 84.4±0.6  & 69.5±1.3  \\\hline
\textbf{Proposed w/o MSP}   & \checkmark         & \checkmark         & \checkmark   &     & 82.4±0.4  & 84.2±0.8  & 70.1±1.1  \\\hline
\textbf{Proposed w/o GRB}   & \checkmark         & \checkmark         &     & \checkmark   & 82.5±0.2  & 83.6±0.4  & 70.8±1.8*  \\\hline
\textbf{Proposed}           & \checkmark         & \checkmark         & \checkmark   & \checkmark   & 83.0±0.2* & 84.6±1.6* & 70.3±1.8 \\\hline
\end{tabular}
\end{table}

\section{Discussion}
\subsection{The limited size of the collected dataset}
In the current study, 122 OCT B-scans from 61 individuals are manually annotated and included in our experiment. While the size of the dataset is relatively small, it should be noted that the qualified human data are difficult to acquire and manually annotating 10 layers in one OCT image is very expensive and time-consuming. To partially overcome this issue, we performed data augmentations on the training set including horizontal flipping, additive Gaussian noises, and contrast adjustment.

\subsection{The impact of imaging artifacts and label noises on the segmentation results}
It is well known that the commonly presented artifacts including vessel shadows and retinal lesions might influence the automated segmentation algorithms. Those artifacts often cause blurred layer boundaries, diminished tissue texture, and altered image contrast, which could potentially lead to a decrease in the segmentation accuracy. 

A good illustration is provided in Fig. 8, where the presented retinal lesion is circled out in the right panel of Fig. 8(a). For conventional deep learning-based algorithms such as U-Net, ReLayNet, and DRUNET, labelling errors are visible as shown in Fig. 8(c)-(e). On the other hand, the proposed method is more robust to this perturbation. We believe this might because these algorithms mainly depend on the texture details of the images to perform the pixel-level classifications, while the proposed method explicitly imposes spatial constraints on to the task which regularizes the task and ensures a better visual outcome in this case as illustrated in Fig. 8(f).

It is also worth mentioning that the proposed method might be affected by the label noises. Due to the limited resources, the manual segmentation of the OCT images is performed collaboratively by two graders under the supervision of a retinal specialist, such that one image is only segmented by one grader. Therefore, it is possible that small label noises are introduced, because the two graders might possess different preferences and styles during the annotation \cite{RN158}. This might slightly impair the performance of the segmentation.

To further address these issues, we plan to perform detection, removal, and inpainting for the artifact regions and tackle the label noise in the future \cite{RN163,RN162}.

\subsection{The proposed framework relies on standardized images}
One of the potential limitations of the proposed framework is that it requires the input images to be well standardized such that all anatomical assumptions or spatial constraints we have made are valid. Take the first stage, the optic disc segmentation network, as an example, it relies on the presumption that the optic disc region and the non-disc regions are arranged in a proper horizontal order as mentioned in Section 2.1. We could be done by perform a registration process prior to segmentation with a goal of registering the optic disc region with a retinal template in the future.

\section{Conclusion}
To address the challenges imposed by the multi-scale features presented in the optic disc and the retinal layers with various thicknesses as well as exploiting the existing anatomical priors, a multi-scale global reasoning module, which is capable of long-range contextual spatial reasoning, is proposed and integrated into a U-Net backbone. Specifically, a two-stage framework is constructed to sequentially segment the optic disc and the retinal layers in peripapillary OCT images. We validated the proposed framework on a collected dataset as well as a public dataset. The experimental results on both datasets showed that the proposed method could considerably improve the segmentation performance of optic nerve head tissues compared with other state-of-the-art techniques. The proposed method achieved 82.0\% and 83.0\% in terms of Dice score and pixel accuracy on average, which is 1.6\% and 2.8\% higher than the performance of ReLayNet. More importantly, the visual quality of the segmented images is greatly enhanced, thanks to the anatomical constraints imposed by the multi-scale global reasoning module. In the future, we will incorporate the proposed segmentation network into the workflow of early-stage glaucoma diagnosis. We also believe the proposed architecture could be domain transferred to other biomedical image segmentation tasks where an abundance of anatomical priors is available. To facilitate the progression of the filed, we make our segmentation dataset as well as the codes available. To our best knowledge, this will be the first public peripapillary retinal OCT dataset.

\section{Acknowledgement}
The computations in this paper were run on the $\pi$ 2.0 cluster supported by the Center for High Performance Computing at Shanghai Jiao Tong University. We sincerely appreciate reviewers for their precious suggestions which help improve this work substantially.

\section{Disclosures}
The authors declare no conflicts of interest.

\bibliography{main}






\end{document}